# Detection of Hard Exudates in Retinal Fundus Images Using Deep Learning

Avula Benzamin*, Chandan Chakraborty*

*School of Medical Science and Technology, Indian Institute of Technology Kharagpur, India

Email: benjamin.avula@iitkgp.ac.in, chandanc@smst.iitkgp.ernet.in

*Abstract*— Diabetic Retinopathy (DR) is a retinal disorder that affects the people having diabetes mellitus for a long time (20 years). DR is one of the main reasons for the preventable blindness all over the world. If not detected early, the patient may progress to severe stages of irreversible blindness. Lack of Ophthalmologists poses a serious problem for the growing diabetes patients. It is advised to develop an automated DR screening system to assist the Ophthalmologist in decision making. Hard exudates develop when DR is present. It is important to detect hard exudates in order to detect DR in an early stage. Research has been done to detect hard exudates using regular image processing techniques and Machine Learning techniques. Here, a deep learning algorithm has been presented in this paper that detects hard exudates in fundus images of the retina.

*Index Terms*—Hard Exudates, Deep Learning, Diabetic retinopathy, Fundus Imaging.

## I. INTRODUCTION

Pancreas secrets insulin to regulate the blood sugar levels in the body. Diabetes mellitus is the disease that occurs when the pancreas does not produce sufficient insulin or when the body parts do not utilize effectively, the insulin produced [15]. Diabetes Mellitus is often referred as Diabetes. It is classified mainly into 2 types. Type 1 diabetes, is known as Insulin dependent diabetes. In type 1 diabetes pancreas does not secrete sufficient insulin. Type 2 diabetes, also known as non-insulin dependent diabetes, is because of ineffective utilization of the produced insulin by the body [2]. According to W.H.O, in the year 2000, out of all the population in the world, 2.8% are having diabetes. This may increase to 4.4% by 2030[1]. Diabetes can cause many sufferings in human body like heart diseases and stroke, kidney failure, amputation of lower limbs and loss of vision [14].

Diabetic Retinopathy (DR) is a retinal vascular disease, which sometimes referred to as ocular manifestation of Diabetes [3][14]. DR is a microvascular complication that comes due to diabetes. DR affects the central vision of the patient and if the disease progress without treatment, patient may get irreversible blindness [5]. DR is one of the significant reason for preventable blindness all over the world. In DR, patient can not perceive any apparent changes in his vision until the disease become worse. But early signs of DR can be detected using fundus imaging. In order to detect DR the retinal fundus image needs to be assessed by the trained ophthalmologist. The bottleneck has been created because of the lack of minimum number of ophthalmologists for the growing DR affected patients hence it was identified that an automated DR screening system is required to assist the ophthalmologist in screening the DR quickly as well as accurately.

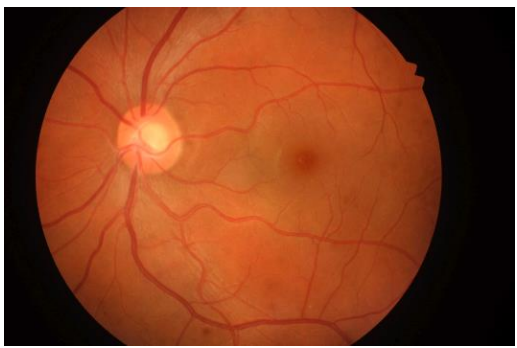 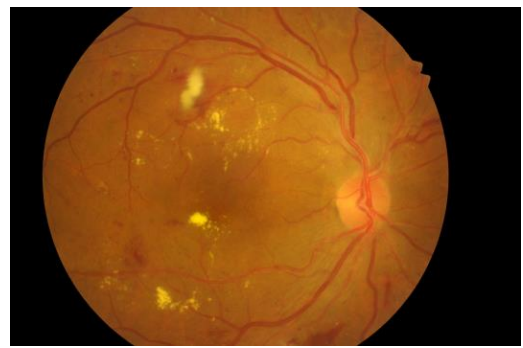

(a) Healthy Retina  (b) Retina with Hard Exudates

Fig 1. Healthy and Unhealthy Retinal fundus images

Microaneurysms are the first pathological symptoms that can be detected using fundus imaging in case of DR. They appear as tiny bulges of blood looks like small red dots on the retina. Microaneurysms causes the blood vessels to leak in retina. As the disease progress lipids and fluids leak from the blood vessels and forms hard exudates that appear in yellowish colour with different shapes and sizes. Macula and fovea are responsible for the central vision, if the exudates formed at these places, patient's central vision may get deteriorated or lost. Detection of the Hard Exudates is very significant in automated detection of DR. A Deep learning algorithm has been developed to detect Hard Exudates. After training, Hard Exudates in the test images are being detected with an accuracy of 98.6% by our deep learning model.

## II. LITERATURE REVIEW

By conducting literature review we have found that Hard Exudates are the most prevailed, most prominent pathological symptoms of DR. detection of DR lesions in fundus images helps to detect DR. Arjun Narang et al [8] implemented an algorithm to detect hard exudates using a Lifting wavelet Transform (LWT) based image enhancement method and Support Vector Machine (SVM) to detect and classify hard exudates. Xian Chen et al [9], using SVM, which is trained on the features extracted from the candidate regions of hard exudates. These candidate regions are selected from fundus images of eye using histogram segmentation with morphological reconstruction [9]. G.G. Rajput et al [10], has worked to detect hard exudates using k-means algorithm. They have used CIELAB color space and preprocessed the images to eliminate noise and, eliminated the blood vessels and then by k-means algorithm they have detected the exudates. Mahdi et al [11], using morphological operations detected hard exudates. They preprocessed the fundus image, eliminated optic disk and retinal blood vessels, then by a mixture of morphological operations like top hat, bottom hat and reconstruction operations hard exudates are segmented. Similar kind of work is done by [13]. Maria Gracia et al [12], have detected exudates using different Machine Learning approaches like Radial basis function, SVM and Multilayer perceptron (MLP).

Most of the work has been done using image processing and Machine Learning approaches. We have decided to work on detection of exudate using deep learning approach.

## III. MATERIALS

We have worked on the retinal fundus images which are labeled for hard exudates. The dataset is downloaded from IDRiD. The fundus images in IDRiD dataset were captured by a retinal specialist at an Eye Clinic located in Nanded, Maharashtra, India. Images were acquired using a Kowa VX-10 alpha digital fundus camera with 50-degree field of view (FOV), and all are centered near to the macula.

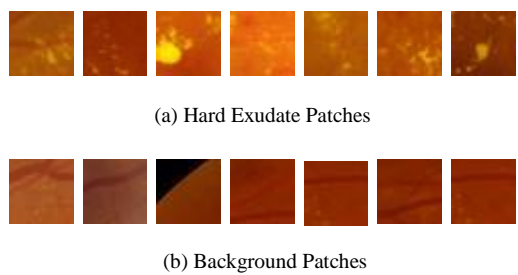

(a) Hard Exudate Patches

(b) Background Patches

Figure 2. Sample Training Patches

## IV. METHOD

Our method is inspired by the work done by Anirban Santara et al [4]. The Dataset is consisted of the retinal fundus images and the corresponding labelled ground truth images for hard exudates. In ground truth images all exudate pixels are having intensity 1, remaining pixels are having intensity 0. We have generated 200000 image patches of dimension 32×32. By using the ground truth we have labelled all patches into two classes, hard exudate and background. 100000 patches belongs to background class, remaining 100000 patches belong to exudate class.

In the extracted image patch of dimension 32×32, the pixel at location (17,17) belongs to either exudate class or background class. The image is labeled as 1 if the interested pixel is hard exudate pixel and as 0 otherwise. Each label is one hot encoded.

The fundus image dataset that consists of 54 images is divided into training set and testing set. Training set is of 40 images and testing set is of 14 images. 2500 hard exudates patches and 2500 background patches were extracted from each image. The network is trained on 200000 image pathces of size 32×32. Sample generated patches of hard exudate class and back ground class are shown in Figure 2.

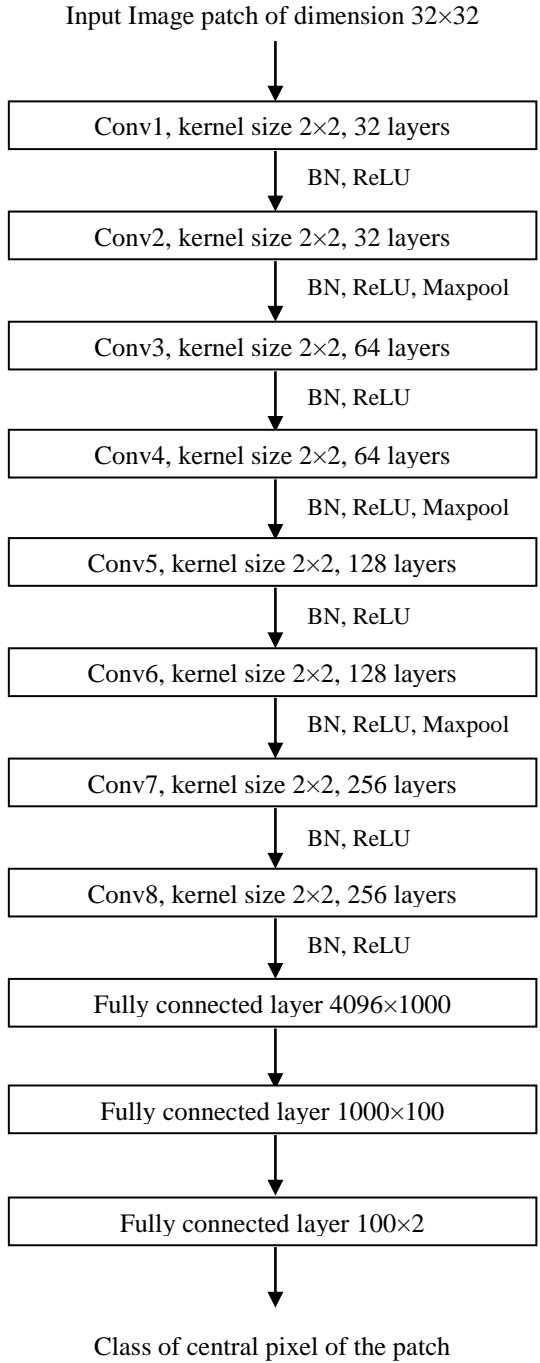

TABLE I. NETWORK ARCHITECTURE

Input Image patch of dimension 32×32
↓
Conv1, kernel size 2×2, 32 layers
BN, ReLU
↓
Conv2, kernel size 2×2, 32 layers
BN, ReLU, Maxpool
↓
Conv3, kernel size 2×2, 64 layers
BN, ReLU
↓
Conv4, kernel size 2×2, 64 layers
BN, ReLU, Maxpool
↓
Conv5, kernel size 2×2, 128 layers
BN, ReLU
↓
Conv6, kernel size 2×2, 128 layers
BN, ReLU, Maxpool
↓
Conv7, kernel size 2×2, 256 layers
BN, ReLU
↓
Conv8, kernel size 2×2, 256 layers
BN, ReLU
↓
Fully connected layer 4096×1000
↓
Fully connected layer 1000×100
↓
Fully connected layer 100×2
↓
Class of central pixel of the patch

All images and ground truths are of same resolution. Each image and its corresponding ground truth in the dataset are resized into a resolution of 256×256 using 'resizeimage' python language module and then the patches were extracted from the training images.

We have developed an 8 layer convolutional neural network of architecture shown in Table 1.

The network predicts whether the central pixel of the patch belongs to either hard exudates class or background class.

The feature map size is halved using maxpool operation after every two convolutional layers, but, at the last convolution layer the feature map size is remained same. Batch normalization is used for faster training [6]. Drop out has been used in order to prevent the network from overfitting [7]. 200000 training images were divided into 5 sets of 40000 images and each set is trained for 500 epochs one after another. We called it a streak when the network trained by all the training images. The network is trained for 3 complete streaks, thus we can say that the network is trained for 1500 epochs. The network achieved 99.4% training accuracy.

Adamoptimizer has been used as optimizer, learning rate is 0.0001, with Cross Entropy as Loss function. Training process is carried out in mini batches. Each mini batch consisted of 50 training images.

The network architecture is presented in Table 1.

After the training, each test image is dissolved into 50176 patches. And each patch is predicted by the network, that the central pixel of the patch is either hard exudate or background. The predictions are reshaped into an image resolution 224×224. We did not used padding while extracting the patches hence all the pixels are not predicted.

The ground truth image also resized to 256×256. And the image is cropped from co-ordinates (16,16) to (240,240) to evaluate the predicted image accuracy. Each test image is predicted with an average accuracy of 98.6%.

Here we are generating an image in which hard exudates are detected by predicting the class of every pixel. We are taking a 32×32 image patch in which the interested pixel is at the location (17,17) of the patch. In this way every pixel in the image is classified by using a 32×32 patch, but, we have not predicted the pixels at the edges of the image.

The resultant images were shown in Figure 2.

## V. RESULTS AND DISCUSSION

The Deep learning model that has been presented in this paper is developed using Tensorflow deep learning framework. The resultant segmented images are shown in figure 3. As we have converted this segmentation problem into a two class classification problem, the sensitivity and specificity of the test image patches are calculated using confusion matrix presented in Table2. The model has predicted all the test image patches with an accuracy of 98.6%. We have considered 100352 patches of two train images for testing. The confusion matrix of the predictions of test images is shown in Table 2. The model predicted the test patches with a Sensitivity of 98.29% and a Specificity of 41.35%. The accuracy of the model over test image patches is 96.6%.

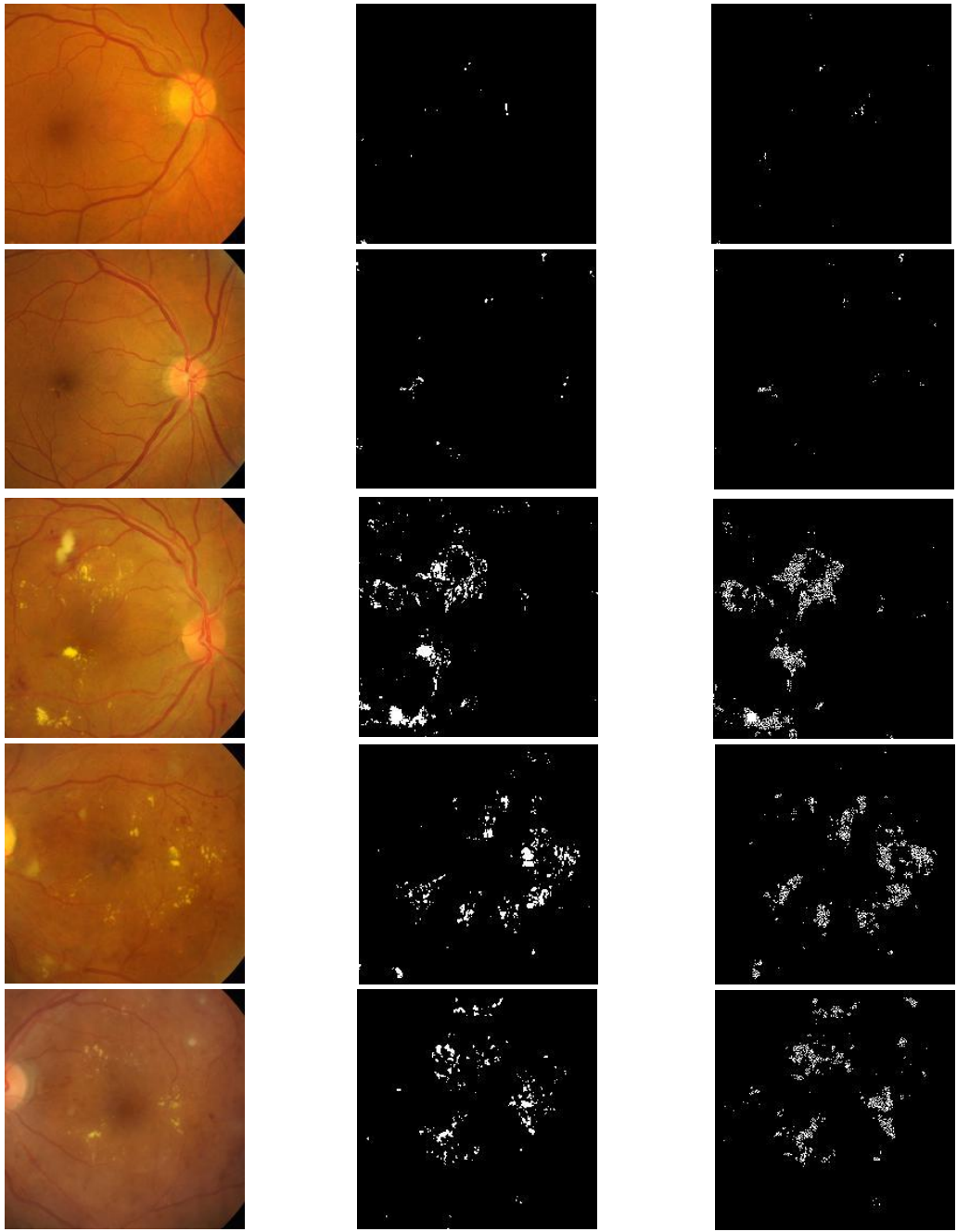

(a) DR affected Fundus Image    (b) Corresponding Ground truth    (c) Predicted Image

Figure 3: Comparison of Original, Ground truth and Predicted Images

## VI. CONCLUSION AND FUTURE SCOPE

Hard Exudates that are present in the DR affected fundus image have been detected using the deep learning model developed in this work. In case of Diabetic retinopathy disease, detection of hard exudates is of interest to detect the presence of disease in an automated way [8]. The presented model detects the hard exudates with an accuracy 98.6

TABLE II. CONFUSION MATRIX

| Predicted / Actual | Background | Exudates | Total |
|---|---|---|---|
| Background | 95862 | 1665 | 97527 |
| Exudates | 1651 | 1174 | 2825 |
| Total | 97513 | 2839 | 100352 |

In future we also want to detect Soft exudates, hemorrhage and microaneurysms. And using the segmented images we want to grade the severity of the presence of DR. We want to increase the accuracy of the model for that we are proposing 2 approaches. First approach is to vary the size of the image patch. Then we can find the effect of image patch on the accuracy of prediction. Second approach is to use an ensemble of convolutional neural networks. As mentioned in section 4, we have not predicted the pixels at the edges of the image i.e., first and last 16 pixels of every row and column were not predicted. We want to overcome this difficulty by zero padding the image before predicting.